\documentclass[12pt]{nature}
\usepackage{amsmath}
\usepackage{amsfonts}
\usepackage{amssymb}
\usepackage{hyperref}
\usepackage{xr-hyper}
\ifdefined\HCode\else
\fi
\usepackage[export]{adjustbox}
\usepackage{graphicx,psfrag,color}% Include figure files
\usepackage{dcolumn}% Align table columns on decimal point
\usepackage{bm}% bold math
\usepackage{appendix}
\usepackage{printlen}
\usepackage{subcaption}
\usepackage[justification=Justified]{caption}
\usepackage{diagbox}
\usepackage{comment}
\usepackage[normalem]{ulem}
\usepackage{float}
\usepackage{overpic}
\newcolumntype{t}[1]{D{.}{.}{#1}}

\newcommand{\sumij}{\sum_{\langle i j \rangle}}
\begin{document}
\title{Pattern Expansion of Spin Glasses} %\\ 

\author{Mutian Shen$^{1,2,\ast}$, Zohar Nussinov$^{1,3,\ast}$, Yang-Yu Liu$^{4,\ast}$}

%\date{\today}
\date{}
\maketitle
\begin{affiliations}
    \item Department of Physics, Washington University, St. Louis, MO 63160, USA
    \item Department of Biomedical Engineering, Washington University, St. Louis, MO 63160, USA
    \item LPTMC, CNRS-UMR 7600, Sorbonne Universit\'e, 4 Place Jussieu, 75252 Paris cedex 05, France
    \item Channing Division of Network Medicine, Department of Medicine, Brigham and Women's Hospital and Harvard Medical School, Boston, Massachusetts 02115, USA
    \\ \noindent $^\ast$Correspondence should be addressed to M.S. (smutian@wustl.edu), Z.N. (zohar@wustl.edu), or Y.-Y.L. (yyl@channing.harvard.edu).
\end{affiliations}

\begin{abstract}
We introduce a systematic method for expanding general spin-glass Hamiltonians in terms of Mattis interactions, providing a novel perspective for understanding the fundamental differences between short-range Edwards-Anderson (EA) and mean-field Sherrington-Kirkpatrick (SK) spin glasses. By iteratively extracting patterns from the coupling matrix, we expand the original spin-glass system into a Hopfield-like model (a series of Mattis interactions) plus a residual system. Our analysis reveals profound distinctions between EA and SK models: while EA models in two and three dimensions break into isolated subconnected sections after expansion, the SK model exhibits remarkable self-similar behavior, with the residual system preserving the mean-field structure and Gaussian statistics throughout the expansion process. This self-similarity manifests in exponential decay of residual matrix norms and expansion coefficients, reflecting the inherent mean-field nature of the SK model. Furthermore, we demonstrate that pattern expansion can identify ultra-low energy excitations in EA models, revealing excitations with energies that decrease rapidly with expansion step. Through connected component analysis, we quantify the size-energy relationship of these independent excitation clusters, opening new avenues for understanding the low-energy landscape of spin glasses and providing insights into the nature of metastable states.
\end{abstract}
\section{Introduction}
Spin glasses, as paradigmatic examples of disordered systems, exhibit simple mathematical formulations yet harbor profound complexity \cite{SG1,Steinbook,Fortunatobook}. Despite decades of intensive investigation, fundamental questions regarding their nature remain unresolved. The Edwards-Anderson (EA) model \cite{EA} describes short-range spin glasses on finite-dimensional lattices, while the Sherrington-Kirkpatrick (SK) model \cite{SK} represents mean-field spin glasses with all-to-all interactions, representing two fundamentally different classes of spin-glass systems. While replica symmetry breaking (RSB) theory \cite{parisi_infinite_1979} provides a rigorous framework for the SK model, its applicability to finite-dimensional EA systems remains a subject of ongoing debate, with alternative pictures such as the droplet theory \cite{mcmillan_scaling_1984,FisherHuse88}, trivial-non-trivial (TNT) scenarios \cite{TNT1}, and chaotic pairs \cite{newman_metastate_1997} offering competing perspectives.

Understanding the low-energy state space of spin glasses is of fundamental importance. Spin-glass systems possess highly complex energy landscapes, with their low-energy regions filled with numerous near-degenerate local minima---the \emph{metastable states}. The organization of these metastable states directly determines the system's dynamical behavior, the difficulty of optimization and sampling algorithms, and when local approximation methods fail or succeed. Therefore, understanding the \emph{structural organization of low-energy states} in spin glasses, rather than just the single ground state, is a central problem. 

There are two main approaches to studying the low-temperature physics and low-energy free-energy landscape of spin glasses. The first is direct Monte Carlo (MC) simulations, which sample the low-energy configurations through thermal fluctuations. The second approach involves computing the system's ground state first, and then perturbing the system using specific methods to probe excited states. This latter approach includes techniques such as domain walls and zero-energy droplets (obtained, for example, by single-bond flipping). These structures frequently appear in numerical simulations, but they do not naturally correspond to simple single-spin or finite-order interactions. A fundamental question arises: Does there exist a mesoscopic description that can systematically characterize the role of these low-energy structures in the energy landscape?

The Hopfield model \cite{hopfield1982neural}, originally introduced in the context of neural networks, can be viewed as a special type of spin glass where the coupling matrix is constructed from stored patterns. The Hopfield model, as a summation of Mattis interaction terms\cite{mattis1976solvable}, has the coupling strength between spins $i$ and $j$ given by $J_{ij} = \sum_{\mu=1}^{\kappa} \alpha_\mu \xi_i^{(\mu)} \xi_j^{(\mu)}$ for $\kappa$ patterns $\boldsymbol{\xi}^{(\mu)} = (\xi_1^{(\mu)}, \xi_2^{(\mu)}, \ldots, \xi_N^{(\mu)})$, where $\xi_i^{(\mu)} = \pm 1$ and $\alpha_\mu$ are the pattern weights. In most cases, $\alpha_\mu$ are constant (typically $\alpha_\mu = 1$). A single term of the form $J_{ij} = \xi_i \xi_j$ for a pattern $\boldsymbol{\xi} = (\xi_1, \xi_2, \ldots, \xi_N)$ is called a \emph{Mattis interaction}. The ground state of such a Mattis interaction system is simply the pattern itself, $\boldsymbol{\sigma}_0 = \boldsymbol{\xi}$, making it analytically tractable. This simplicity suggests that Mattis interactions, and by extension the Hopfield model, might serve as a natural basis for decomposing and understanding more complex spin-glass systems.

In this work, we introduce a systematic method for expanding general spin-glass Hamiltonians in terms of Mattis interactions. This \emph{pattern expansion} provides a novel perspective for understanding the fundamental differences between short-range (EA) and mean-field (SK) spin glasses. The key idea is to expand the complex energy landscape of a spin glass into simpler energy landscapes corresponding to Mattis interactions, whose energy landscapes are in fact equivalent to those of ferromagnetic models. By iteratively extracting patterns from the coupling matrix, we expand the original spin-glass system into a series of Mattis interactions plus a residual system. The effective energy of the system can then be expressed as a superposition of these Mattis interactions, with generally unequal weights reflecting their different importance in the energy landscape. This representation is structurally similar to the \emph{Hopfield model}, but our focus is not on ``memory'' or ``storage,'' but rather on viewing the Hopfield form as an \emph{effective expansion of the spin-glass energy landscape}.

The rationale for this decomposition is twofold. First, the expansion terms—the “free” part of the system—are analytically transparent, as they consist of simple Mattis interaction terms closely related to the Hopfield model. Each term defines a tractable energy landscape with a well-defined ground state, and together they form a structural backbone of the full system. Second, the residual system—the “perturbative” part—encodes the corrections and interactions beyond this Mattis-based approximation. This separation is conceptually analogous to perturbative constructions in quantum field theory\cite{peskin2018introduction,srednicki2007quantum}, where physical observables are formulated in the space spanned by eigenstates of a free theory, with interactions treated separately. Importantly, this analogy is purely heuristic: pattern expansion does not define a controlled perturbative expansion, but rather provides a framework for isolating and interpreting the role of the residual system as the source of nontrivial many-body interactions beyond the Mattis backbone.
%This analogy with quantum field theory does not mean that we adopt the entire procedure of perturbation theory, but rather emphasizes the potential importance of this paradigm. 
Remarkably, we find that the behavior of this residual system after multiple expansion steps reveals profound distinctions between EA and SK models.

\begin{figure}[htb]
    \centering
    \begin{overpic}[width=0.44\textwidth]{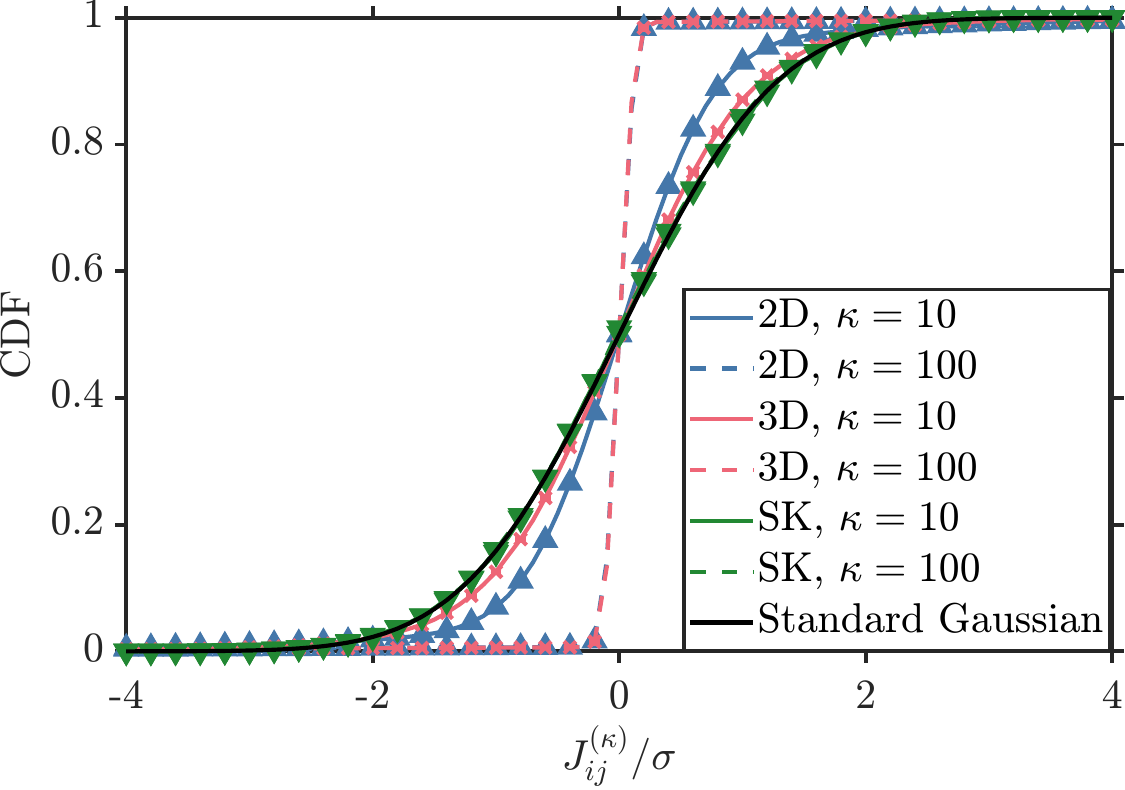}
        \put(4,74){\fontsize{12}{16}\fontfamily{phv}\selectfont{\textbf{a}}}
    \end{overpic}%
    \hspace{0.02\textwidth}%
    \begin{overpic}[width=0.46\textwidth]{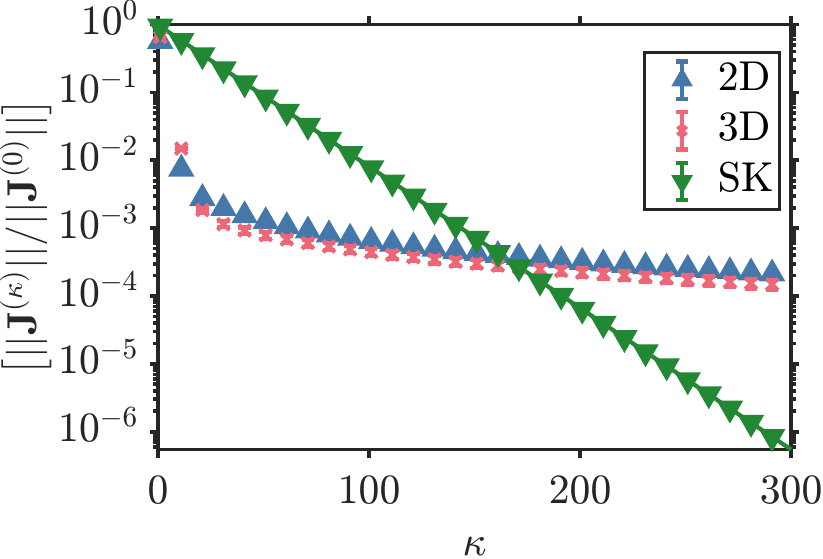}
        \put(8,70){\fontsize{12}{16}\fontfamily{phv}\selectfont{\textbf{b}}}
    \end{overpic}
    \caption{\textbf{Residual coupling statistics and norms under pattern expansion in EA and SK spin glasses.} (a) Cumulative distribution function (CDF) of normalized residual coupling matrix elements $J_{ij}^{(\kappa)} / \sigma^{(\kappa)}$ for $\kappa = 10$ (solid lines) and $\kappa = 100$ (dashed lines) in 2D EA (blue), 3D EA (orange), and SK (green) models. The black line shows the standard Gaussian CDF for comparison. For the SK model, the CDF remains close to Gaussian at all expansion steps, demonstrating self-similarity: the residual system preserves the statistical properties of the original SK model. In contrast, EA models deviate from Gaussian behavior, particularly at larger $\kappa$ values, indicating the breakdown of self-similarity. (b) Normalized residual matrix norm $\left[ \|\mathbf{J}^{(\kappa)}\|/\|\mathbf{J}^{(0)}\| \right]$ as a function of expansion step $\kappa$ for 2D EA (triangles), 3D EA (crosses), and SK (inverted triangles) models. For SK model, quantities are normalized by $\sqrt{N}$ to account for mean-field scaling. SK shows near-exponential decay, while EA decay slows down as the system fractures into isolated islands. Additional details are shown in Supplementary Fig.~\ref{fig:si_matrix_norm_alphas}. Error bars represent standard errors across multiple samples.}
    \label{fig:matrix_norm_alphas}
\end{figure}

These findings focus on two perspectives: first, the characterization of the residual system's intrinsic properties, and second, understanding the impact of the residual system on the entire system (excitations). From the first perspective, we observe that the residual system behaviors of lattice (EA) spin glasses and mean-field (SK) spin glasses are fundamentally different. For EA models in two and three dimensions, the residual system after expansion typically consists of isolated subconnected sections (see Fig.~\ref{fig:decomp_illu}(a)--(b) and Supplementary Fig.~\ref{fig:si_normalized_energy_components}). In contrast, the SK model exhibits a pronounced self-similar behavior: the residual system after each expansion step remains another SK model, preserving the mean-field structure throughout the expansion process (see Fig.~\ref{fig:decomp_illu}(c) and Fig.~\ref{fig:matrix_norm_alphas}(a)). This self-similarity manifests itself in several ways: (i) the probability distribution of residual coupling elements remains Gaussian at all expansion steps (Fig.~\ref{fig:matrix_norm_alphas}(a)), (ii) the residual matrix norm decays exponentially (Fig.~\ref{fig:matrix_norm_alphas}(b)), and (iii) the expansion coefficients $\alpha_\kappa$ follow an exponential decay (Supplementary Fig.~\ref{fig:si_matrix_norm_alphas}). This self-similarity, which we characterize both numerically and analytically, reflects the inherent mean-field nature of the SK model and distinguishes it from finite-dimensional systems. 

Second, we demonstrate that pattern expansion can be employed to identify ultra-low energy excitations in EA models. By analyzing the cumulative expansion states $\boldsymbol{\sigma}^{(\kappa)}$ and their relationship to the original ground state, we uncover excitations with energies $\Delta E$ that decrease rapidly with expansion step $\kappa$ (Fig.~\ref{fig:energy_spin_differences}(a) and (c)). Through connected component analysis, we identify independent excitation clusters and quantify their size-energy relationship (Fig.~\ref{fig:energy_spin_differences}(b) and (d)). These excitations, which are difficult to detect through conventional methods, open new avenues for understanding the low-energy landscape of spin glasses and may provide insights into the nature of metastable states.

\section{Formalism}\label{sec:formalism}
A general form of the Ising spin glass model is given by:
\begin{eqnarray}
\label{HSG}
    \mathcal{H} = - \sum_{\langle i j \rangle} J_{ij} \sigma_i\sigma_j.
\end{eqnarray}%conditions
A \emph{Mattis interaction} is a special type of spin-glass coupling where the coupling strength $J_{ij}$ is determined by a single pattern $\boldsymbol{\xi}=(\xi_1,\xi_2,\dots,\xi_N)$, where $\xi_i=\pm 1$ and $N$ is the number of spins:
\begin{equation}
    \label{eq:hopfield}
    J_{ij}^{\rm Mat}=\Xi^{(\xi)}_{ij} \equiv \xi_i\xi_j.
\end{equation}
The ground state of such a Mattis interaction system, $\boldsymbol{\sigma}_0$, that minimizes the system energy (Eq.(\ref{HSG})) is the same as the pattern $\boldsymbol{\xi}$, $\boldsymbol{\sigma}_0=\boldsymbol{\xi}$. The Hopfield model is then a summation of multiple Mattis interaction terms: $J_{ij}^{\rm Hop} = \sum_{\mu=1}^{\kappa} \alpha_\mu \xi_i^{(\mu)} \xi_j^{(\mu)}$.

We want to expand the Hamiltonian of spin glasses (Eq. \eqref{HSG}) in terms of Mattis interactions (Eq. \eqref{eq:hopfield}), namely we expect:
\begin{equation}
    \label{eq:expansion}
    J_{ij}\approx \sum_{\kappa=1}^{M}\alpha_\kappa \Xi^{(\kappa)}_{ij}.
\end{equation}
That is, we want to minimize $\| J_{ij}-\sum_{\kappa=1}^{M}\alpha_\kappa \Xi_{ij}^{(\kappa)}\|_2$. One way to do that is `step-by-step expansion', just like how Taylor expansion works. Firstly, we minimize 
\begin{align}
    &\sum_{\langle i j \rangle}(J_{ij}-\alpha_1 \Xi_{ij}^{(1)})^2 = \sum_{\langle i j \rangle}J^2_{ij}-2\sum_{\langle i j \rangle}J_{ij}\Xi_{ij}^{(1)}\alpha_1+|\mathcal{E}|\alpha_1^2 \nonumber\\
    &=\sum_{\langle i j \rangle}J^2_{ij}+|\mathcal{E}|(\alpha_1-\frac{\sum_{\langle ij \rangle}J_{ij}\Xi_{ij}^{(1)}}{|\mathcal{E}|})^2 -\frac{(\sum_{\langle ij \rangle} J_{ij}\Xi_{ij}^{(1)})^2}{|\mathcal{E}|},
\label{eq:minimize}
\end{align}
where $|\mathcal{E}|$ is the number of bonds. The minimization requires $\alpha_1 = \frac{\sum_{\langle ij \rangle}J_{ij}\Xi_{ij}^{(1)}}{|\mathcal{E}|}$ and $\boldsymbol{\xi}^{(1)}$ being the ground state of either $\mathbf{J}$ or $-\mathbf{J}$. It is necessary to point out here that, for EA models, the ground states of $\mathbf{J}$ and $-\mathbf{J}$ are actually identical - this can be easily proven through gauge symmetry by a checkerboard style gauge transformation (see Ref.\cite{SM}, Sec.~\ref{sec:sm_gauge}).
 Then $J_{ij} \to J^{(1)}_{ij}=J_{ij}-\alpha_1 \Xi^{(1)}_{ij}\equiv J_{ij}-\frac{\sum_{\langle lm \rangle}J_{lm}\Xi_{lm}^{(1)}}{|\mathcal{E}|}\Xi_{ij}^{(1)}$. We can do it again and again, such that:
\begin{equation}
    J^{(\kappa)}_{ij} = J^{(\kappa-1)}_{ij}-\alpha_{\kappa}\Xi^{(\kappa)}_{ij}.
\end{equation}
For consistency, we can define $J^{(0)}_{ij}\equiv J_{ij}$. Here, $J_{ij}^{(\kappa)}$ denotes the residual system after $\kappa$ steps of pattern expansion. In general, we have:
\begin{equation}
    J_{ij} = \sum_{\mu=1}^\kappa \alpha_\mu \Xi^{(\mu)}_{ij}+J^{(\kappa)}_{ij}.
\end{equation}
The cumulative expansion, up to step $\kappa$, defines a partial reconstruction of the original coupling matrix: $\mathbf{J}_{\text{cumu}}^{(\kappa)} = \sum_{\mu=1}^{\kappa} \alpha_\mu \mathbf{\Xi}^{(\mu)}$, where $\mathbf{\Xi}^{(\mu)}_{ij} = \xi_i^{(\mu)} \xi_j^{(\mu)}$ corresponds to the extracted pattern $\boldsymbol{\xi}^{(\mu)}$ at step $\mu$. The ground state of this cumulative Hamiltonian, denoted as $\boldsymbol{\sigma}^{(\kappa)} = \text{GS}(\mathbf{J}_{\text{cumu}}^{(\kappa)})$, represents an approximation to the original system's ground state and serves as a low-energy excitation when compared to the true ground state $\boldsymbol{\sigma}_0$ of the original system.

Having introduced the method of pattern expansion, we now apply this technique to both SK and EA models. Our analysis focuses on two main aspects. First, what properties does the residual system exhibit after several expansion steps? Second, how much information about the original system is captured by the sum of the first several expansion terms, which serves as an approximation to the original system? 

As preliminary illustrations, Fig.~\ref{fig:decomp_illu}(a)--(c) and (d) correspond to these two aspects, respectively. Panels (a)--(c) not only show the expansion process for different spin-glass systems (2D EA, 3D EA, and SK models), but also allow us to observe the distribution properties of the residual system through the final residual term, providing an intuitive demonstration of the fundamental differences in residual system properties between EA and SK models during the expansion process. Panel (d) shows a schematic illustration indicating that the complex energy landscape of a spin glass can be expanded into a sum of simpler energy landscapes through pattern expansion. Remarkably, only a small number of the first few expansion terms may be sufficient to provide a good approximation to the original system. As we will discuss later, this is indeed the case for 2D and 3D EA spin glasses.

\section{Self-Similarity of Sherrington-Kirkpatrick Model and Finite Expansion of Edwards-Anderson Models}\label{sec:sk_similarity}

We examine the residual system from two complementary perspectives. First, we analyze the distribution of coupling strengths $J_{ij}$ in the residual system. Second, we examine the size of the residual system, quantified by the matrix norm $\sumij {J^{(\kappa)}_{ij}}^2$. The SK and EA models exhibit markedly different characteristics in both regards after pattern expansion. 

For EA models, the weight distribution rapidly concentrates toward zero, meaning that ``non-trivial'' bond connections become rare, see Fig.~\ref{fig:matrix_norm_alphas}(a). In fact, as shown in Supplementary Fig.~\ref{fig:si_normalized_energy_components}, for the sparse residual system structure, bonds mostly exist as small clusters. On the other hand, we can observe in Fig.~\ref{fig:matrix_norm_alphas}(b) that the norm of the residual system decays rapidly in the first few expansion steps---this suggests that the first few expansion terms are sufficient to provide a good approximation to the entire system. 

\begin{figure}[htb]
    \centering
    \begin{overpic}[width=0.99\textwidth]{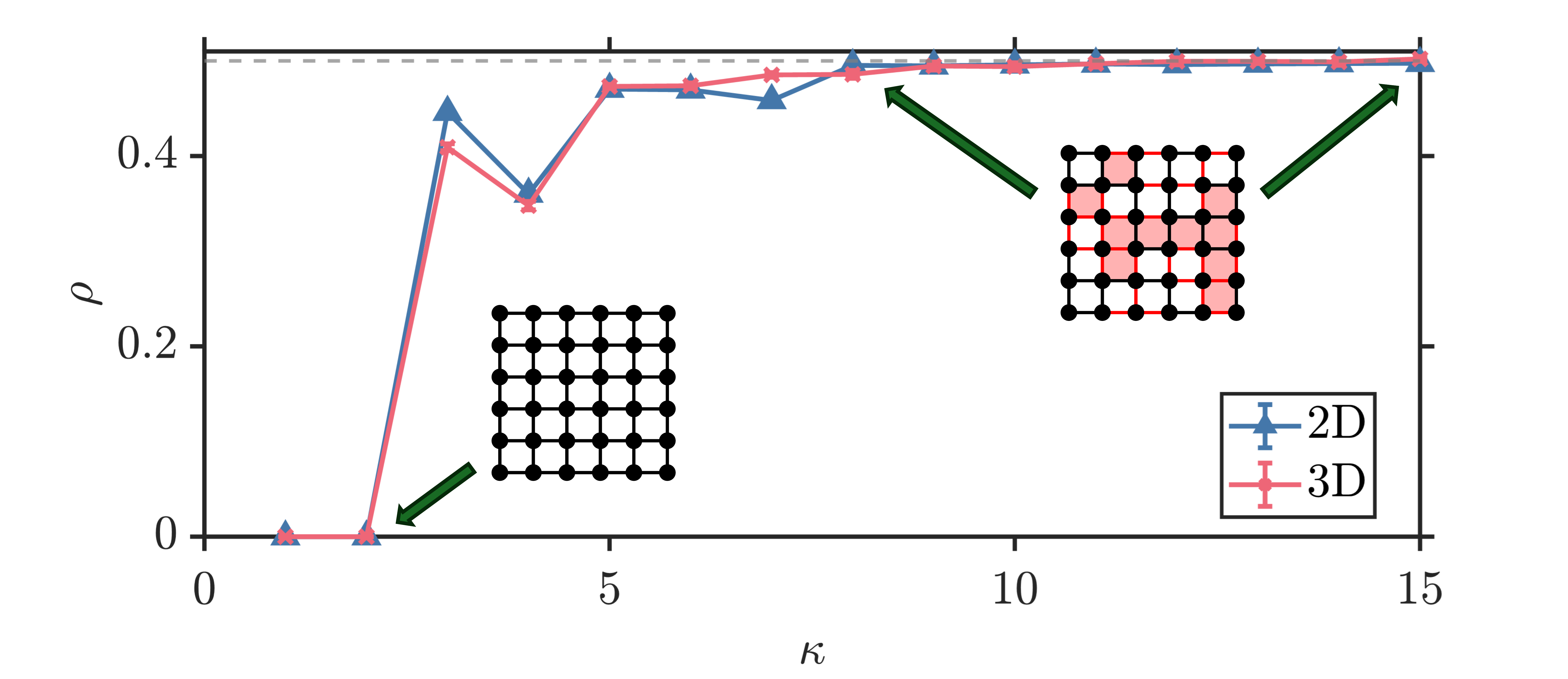}
        \put(6,38){\fontsize{12}{16}\fontfamily{phv}\selectfont{\textbf{a}}}
    \end{overpic}
    \hspace*{-0.05\textwidth}%
    \begin{overpic}[width=0.88\textwidth]{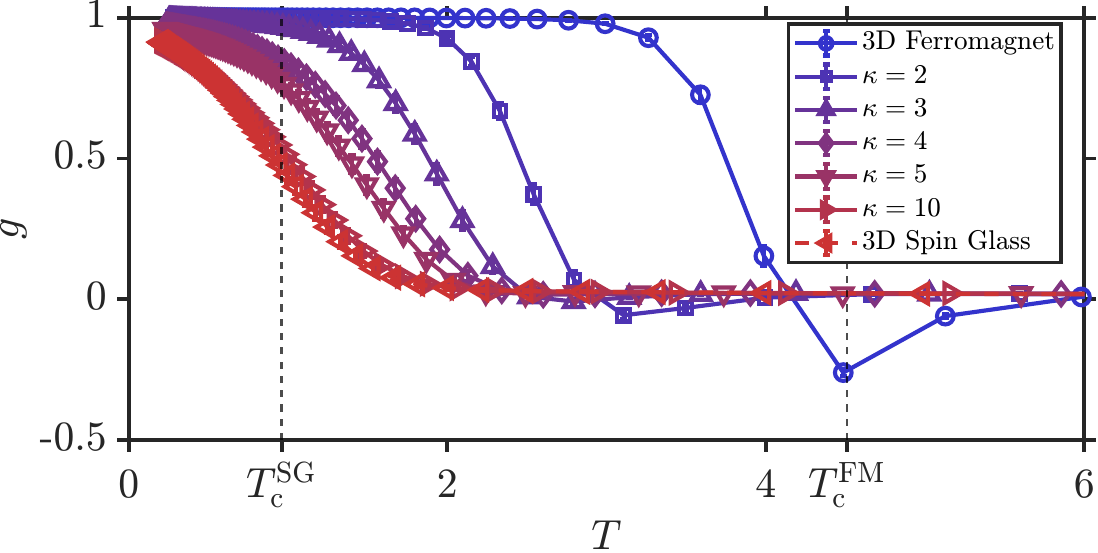}
        \put(3,48){\fontsize{12}{16}\fontfamily{phv}\selectfont{\textbf{b}}}
    \end{overpic}
    \caption{\small \textbf{Frustration density and thermodynamic behavior under pattern expansion in three-dimensional EA spin glasses.} (a) At different expansion orders $\kappa$, the proportion of frustrated plaquettes $\rho$ is statistically analyzed. This quantity serves as an indicator of the ``roughness'' of the energy landscape. For typical spin glasses, the proportion of frustrated plaquettes is close to 0.5. (b) Binder cumulant $g$ of the spin overlap versus temperature $T$ for three-dimensional Edwards--Anderson systems with $L=4$, computed by population annealing with thermal boundary conditions (Supplementary Sec.~\ref{sec:supp_pa}), for the original coupling matrix and for cumulative matrices at $\kappa=1,2,3,4,5,10$. All coupling matrices are normalized so that the standard deviation of bond strengths $J_{ij}$ (over nonzero couplings) equals unity. Vertical lines mark $T_{\mathrm{c}}$ of the three-dimensional ferromagnetic Ising model $T_{\mathrm{c}}^{\mathrm{FM}}\approx 4.511$\cite{ferrenberg1991critical} and of the three-dimensional Edwards--Anderson spin glass $T_{\mathrm{c}}^{\mathrm{SG}}\approx 0.96$\cite{wangPopulationAnnealingMonte2015}. For the ferromagnet, the inflection near $T_{\mathrm{c}}^{\mathrm{FM}}$ that drives $g$ below zero is also seen in similar calculations\cite{lundowIsingSpinGlasses2017}.}
    \label{fig:frustration}
\end{figure}

From a simple structural perspective, Fig.~\ref{fig:frustration}(a) summarizes how the density of frustrated plaquettes $\rho$ depends on the expansion order $\kappa$. The density of frustrated plaquettes is a convenient characterization of the roughness of the energy landscape; for instance, Ising models with tunable hardness have been constructed by adjusting $\rho$\cite{wang_patch-planting_2017}. In this sense, we find that $\rho$ is already close to that of a zero-mean Gaussian spin glass when $\kappa$ is still in the single-digit regime, so that the first few expansion terms largely set the skeleton of the system.

To address thermodynamic behavior, we study the Binder cumulant of the spin overlap\cite{wangPopulationAnnealingMonte2015}, which is a commonly used order parameter in studies of the spin-glass phase transition,
\begin{equation}\label{eq:binder_cumulant}
    g = \frac{1}{2}\left(3 - \frac{\langle q^4\rangle}{\langle q^2\rangle^2}\right),
\end{equation}
built from the second and fourth moments of the overlap between independent replicas. For orientation, the critical temperature of the three-dimensional ferromagnetic Ising model is $T_{\mathrm{c}}^{\mathrm{FM}}\approx 4.511$\cite{ferrenberg1991critical}, whereas that of the three-dimensional Edwards--Anderson spin glass is $T_{\mathrm{c}}^{\mathrm{SG}}\approx 0.96$\cite{wangPopulationAnnealingMonte2015}. Using population annealing with thermal boundary conditions, as detailed in Supplementary Sec.~\ref{sec:supp_pa}, we obtain $g$ versus $T$ for cubic systems with linear size $L=4$, comparing the original couplings to cumulative matrices at expansion orders $\kappa=1,2,3,4,5,10$ (Fig.~\ref{fig:frustration}(b)), where the $\kappa=1$ cumulative system is essentially ferromagnetic. As shown in Fig.~\ref{fig:frustration}(b), already at $\kappa=10$ the transition curve is very close to that of the spin glass, indicating that assembling only a modest number of patterns can nevertheless reproduce the thermodynamic behavior of a spin glass to good accuracy.

For the SK model, we observe that regardless of how large the expansion order $\kappa$ is (even for $\kappa=100$ or more), the distribution remains almost indistinguishably close to a Gaussian distribution. This seems to indicate that no matter how many expansion steps are performed, the residual system of the SK model remains an SK model. Of course, we recognize that merely examining the weight distribution cannot guarantee this. If the SK model exhibits self-similarity, then we should have $\sumij {J^{(\kappa)}_{ij}}^2 \simeq (1-2e_0^2/N)^\kappa N/2$, suggesting an exponential decay of the residual norm. Using the same approach, we can also show that $\alpha_\kappa$ decays with a factor of $(1-4e_0^2/N)^{\kappa/2}$ (see Ref.\cite{SM}, Sec.~\ref{sec:sm_decay}). As further verification, we compute the normalized residual system norm (Fig.~\ref{fig:matrix_norm_alphas}(b)), as well as the expansion coefficients $\alpha_\kappa$ and energy density over multiple disorder realizations (Supplementary Fig.~\ref{fig:si_matrix_norm_alphas}). We further note that when $N$ is sufficiently large, $\alpha_\kappa$ hardly decays, as $\lim_{N \to \infty} (1-4e_0^2/N)=1$. In this case, even to approximate a SK model realization purely in the sense of matrix norm, we need on the order of $N$ patterns, because $\lim_{N \to \infty} (1-4e_0^2/N)^N = \exp(-4e_0^2)$.

The preservation of Gaussian statistics in the SK model's residual system at each expansion step strongly suggests its mean-field nature. For EA models, connected component analysis of the residual system reveals that after sufficient expansion steps, the residual system consists of isolated disconnected sections---typically small clusters of a few spins.  The numerous small clusters in the sparse residual system of EA models suggest the localized nature of their excitations. In the next section, we will examine these excitations more directly and in greater detail.

\section{Ultra-Low Energy Excitations of Edwards-Anderson Model}\label{sec:ultra_low}

In the study of spin glasses, particularly their low-temperature physics, there exists a traditional paradigm. One first computes the ground state of the system, then perturbs the system in some way (for domain walls\cite{melchert_scaling_2009,fractal2d}, by reversing boundary conditions; for zero-energy droplets (ZED)\cite{shen2023universal}, by setting a bond to its critical threshold) and computes the ground state of the perturbed system. The ground state of the perturbed system can be viewed as an excited state of the original system---this provides a systematic and operational approach to studying low-energy excitations. However, a difficult question remains: to what extent do these systematically constructed low-energy excitations represent, or capture, the truly existing low-energy excitations that are possible in the system?

In some sense, a more natural excitation should be obtained by perturbing the entire system (rather than a single bond or boundary). Methods such as random perturbations have been attempted. In the context of our pattern expansion method, we can naturally view the original system $\mathbf{J}$ as a perturbation of its ``free'' part, the cumulative expansion $\mathbf{J}_{\text{cumu}}^{(\kappa)} = \sum_{\mu=1}^{\kappa} \alpha_\mu \mathbf{\Xi}^{(\mu)}$. In practice, we reverse this perspective: we treat the ground state of $\mathbf{J}_{\text{cumu}}^{(\kappa)}$ as an excited state of $\mathbf{J}$. Specifically, for each expansion step $\kappa$, we compute the cumulative ground state
\begin{equation}\label{eq:cumulative_ground_state}
    \boldsymbol{\sigma}^{(\kappa)} = \text{GS}\!\left(\sum_{\mu=1}^{\kappa} \alpha_\mu \mathbf{\Xi}^{(\mu)}\right) = \text{GS}(\mathbf{J}_{\text{cumu}}^{(\kappa)}),
\end{equation}
where $\text{GS}(\cdot)$ denotes the ground state of the corresponding Hamiltonian.

The key difference between this excitation and those from domain walls or droplets is that, due to the global nature of the perturbation, the resulting excitation is not ``independent'' in the sense that this global excitation consists of multiple disconnected elementary excitations. To analyze this, we compute the energy difference $\Delta E^{(\kappa)} = E(\boldsymbol{\sigma}^{(\kappa)}) - E(\boldsymbol{\sigma}_0)$ and the spin difference $\mathcal{V}^{(\kappa)} = \min\left(|\{i: \sigma_i^{(\kappa)} \neq \sigma_{i,0}\}|, N - |\{i: \sigma_i^{(\kappa)} \neq \sigma_{i,0}\}|\right)$, where $\boldsymbol{\sigma}_0$ is the reference ground state of the original system. We analyze these quantities both at the global level and at the level of individual elementary excitations, as shown in Fig.~\ref{fig:energy_spin_differences}(a) and (c). As the expansion order $\kappa$ increases, both the total energy and volume (spin difference) decrease. In Supplementary Fig.~\ref{fig:si_kappa_cluster_counts}, we further show how the number of elementary excitations varies with expansion step $\kappa$.

To identify the elementary excitations, we employ a connected component analysis. Returning to the traditional paradigm for studying excitations, which considers the energy and size of elementary excitations, we plot the energy and size of all elementary excitations across multiple disorder realizations in Fig.~\ref{fig:energy_spin_differences}(b) and (d). First, we note that most excitations are small in size---this is consistent with the sparse structure of the residual system we analyzed earlier. Second, we observe little correlation between energy and size; in fact, we compute the Pearson correlation coefficient and find that these correlation coefficients are all close to zero (specific values are shown in the figure). This low-energy picture---concentrated size distribution at small sizes and absence of correlation between energy and size---is consistent with what zero energy droplets reveals\cite{shen2023universal}.

\section{Discussion}\label{sec:discussion}

In this work, we have introduced pattern expansion as a systematic method for decomposing spin-glass Hamiltonians into a series of Mattis interactions plus a residual system. This approach provides a novel perspective for understanding the fundamental differences between short-range (EA) and mean-field (SK) spin glasses, revealing deep structural distinctions that are not immediately apparent from conventional analyses. The pattern expansion method offers several advantages over conventional approaches. First, it provides a systematic, hierarchical decomposition of the energy landscape, where each expansion step extracts the most important pattern at that level. Second, it naturally identifies low-energy excitations without requiring ad hoc perturbations or domain wall constructions. Third, it reveals structural differences between EA and SK models that are not apparent from conventional analyses. Finally, the residual system itself provides insights into the system's fundamental properties, analogous to how perturbative corrections in quantum field theory reveal interaction structures.

Our key finding is the remarkable self-similarity exhibited by the SK model under pattern expansion. The residual system at each expansion step preserves the Gaussian statistics and mean-field structure of the original SK model, with exponential decay of residual matrix norms and expansion coefficients. This self-similarity directly reflects the inherent mean-field nature of the SK model, where the all-to-all connectivity ensures that removing any finite number of patterns cannot fundamentally alter the statistical structure of the system. In contrast, EA models break down into isolated subconnected sections after sufficient expansion steps, with the residual system consisting of small clusters of spins. This fundamental difference highlights the distinct nature of short-range versus mean-field spin glasses at a structural level.

The self-similarity of the SK model under pattern expansion provides a new perspective on the RSB picture\cite{parisi1979infinite,parisi2020theory}. The fact that the residual system remains an SK model at all expansion steps suggests that the complexity of the SK energy landscape is not captured by a finite number of dominant patterns, but rather emerges from the collective behavior of an infinite hierarchy of patterns. This is consistent with the RSB framework, where the energy landscape is characterized by an ultrametric structure with infinitely many levels of hierarchy. 

\begin{figure}[htb]
    \centering
    \begin{overpic}[width=0.51\textwidth]{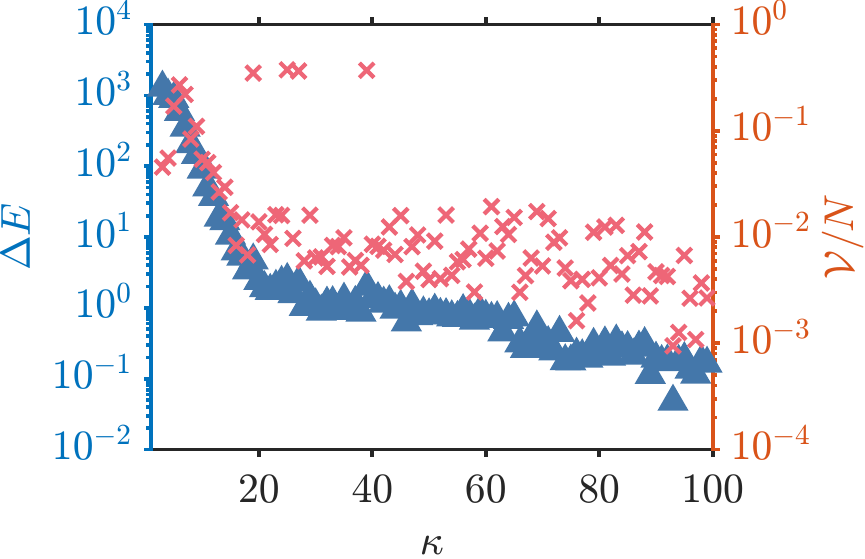}
        \put(5,68){\fontsize{12}{16}\fontfamily{phv}\selectfont{\textbf{a}}}
        \put(-10,36){\fontsize{12}{16}\fontfamily{phv}\selectfont{\textbf{2D}}}
    \end{overpic}%
    \begin{overpic}[width=0.42\textwidth]{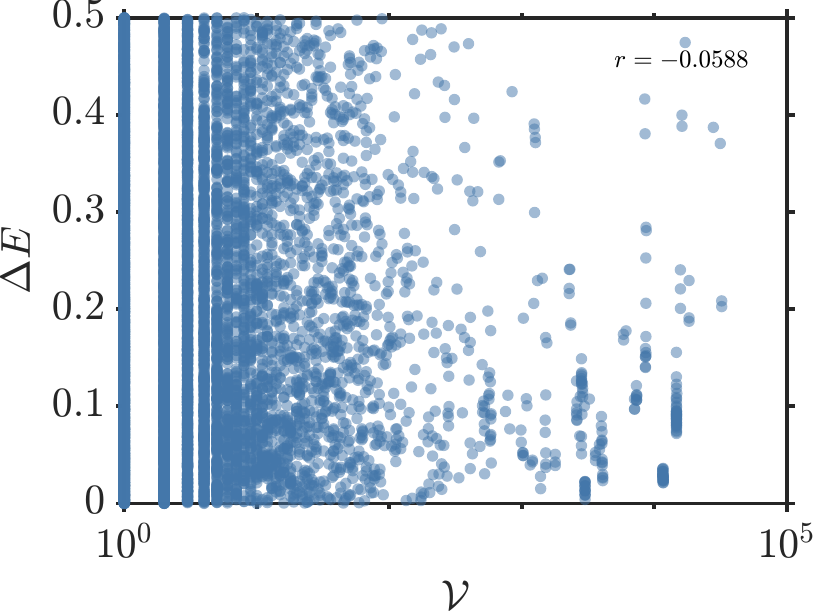}
        \put(2,81){\fontsize{12}{16}\fontfamily{phv}\selectfont{\textbf{b}}}
    \end{overpic}%
    \\
    \begin{overpic}[width=0.51\textwidth]{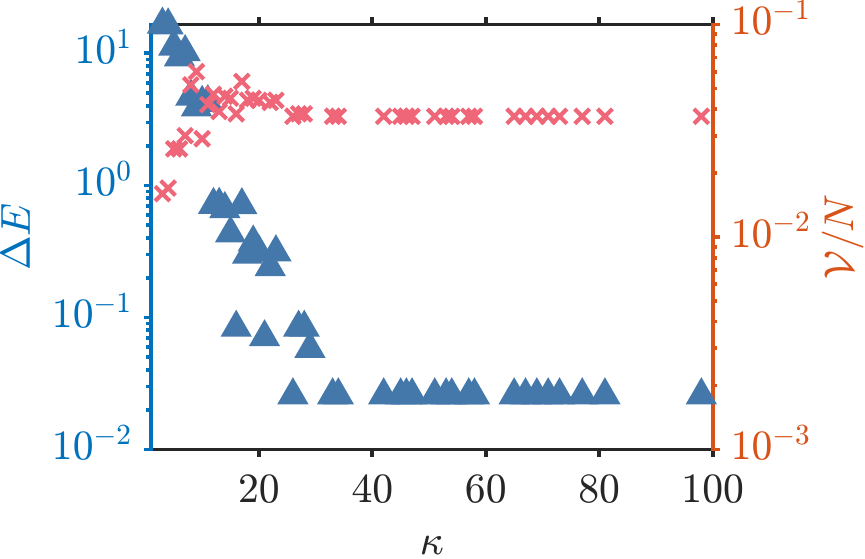}
        \put(5,68){\fontsize{12}{16}\fontfamily{phv}\selectfont{\textbf{c}}}
        \put(-10,36){\fontsize{12}{16}\fontfamily{phv}\selectfont{\textbf{3D}}}
    \end{overpic}
    \begin{overpic}[width=0.42\textwidth]{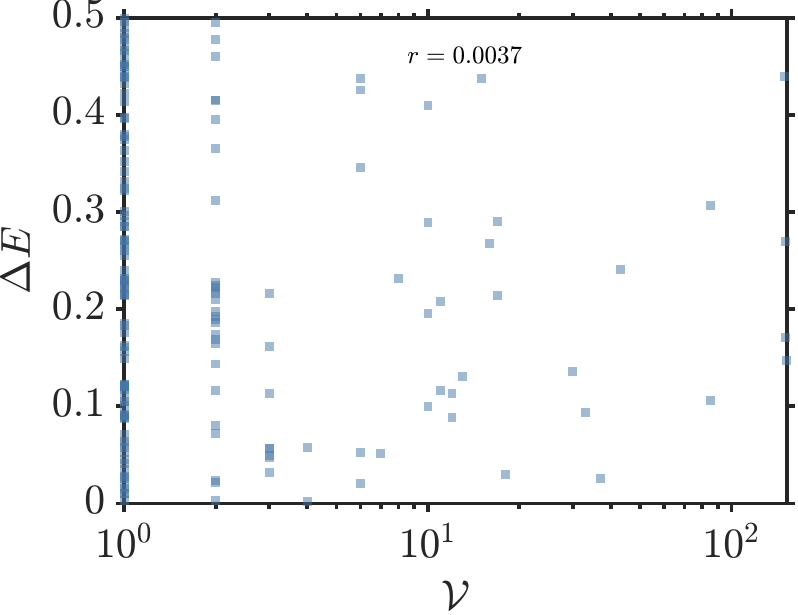}
        \put(2,81){\fontsize{12}{16}\fontfamily{phv}\selectfont{\textbf{d}}}
    \end{overpic}
    \caption{\small \textbf{Ultra-low energy excitations revealed by pattern expansion in two- and three-dimensional EA spin glasses.} In two-dimensional spin glasses, the energy changes caused by single-spin flips are analyzed. The \emph{spin difference} is defined as a measure of local energy response. Panels (a) and (b) show 2D illustrations, while (c) and (d) show 3D generalizations; the method is consistent, with only the spatial dimension differing. (a) and (c) show energy and spin differences as functions of expansion step $\kappa$ for 2D EA (a) and 3D EA (c) models. The left y-axis shows the energy difference $\Delta E^{(\kappa)}$, while the right y-axis shows the normalized spin difference $\mathcal{V}^{(\kappa)}/N$. For 2D and 3D cases, multiple excitations often exist, so $\Delta E^{(\kappa)}$ and $\mathcal{V}^{(\kappa)}/N$ are computed in the sense of summing over all excitations. (b) and (d) show scatter plots of ``energy difference vs. size'' for independent excitation clusters in 2D (b) and 3D (d) EA models. Each point corresponds to a connected excitation cluster, with the horizontal axis showing the normalized size $\mathcal{V}/N$ and the vertical axis showing the cluster energy difference $\Delta E_{\text{cluster}}$. The correlation coefficient is also shown in the figure to characterize whether there is a significant correlation between energy and size.}
    \label{fig:energy_spin_differences}
\end{figure}

For EA models, both the residual system analysis and excitation analysis reveal that ultra-low energy excitations are localized. Firstly, the residual system breaks down into isolated small clusters after sufficient expansion steps. Secondly, the ability of pattern expansion to identify ultra-low energy excitations reveals that the cumulative expansion states $\boldsymbol{\sigma}^{(\kappa)}$ represent systematically constructed low-energy configurations, whose energies decrease rapidly with expansion step $\kappa$. The connected component analysis shows that these excitations form independent clusters, each with its own energy cost, and these clusters are typically small---just a few spins. However, our analysis reveals that there is no significant correlation between the energy and size of these excitation clusters (see Fig.~\ref{fig:energy_spin_differences}(b) and (d)). This finding does not fully support the droplet picture\cite{mcmillan_scaling_1984,FisherHuse88}, which predicts that low-energy excitations should exhibit a scaling relationship between energy and size.

There are, however, several limitations and open questions. While we have demonstrated the method for 2D and 3D EA models and the SK model, its applicability to other spin-glass systems (such as long-range models or models with different interaction distributions) remains to be explored. The relationship between the extracted patterns and physical observables (such as correlation functions or overlap distributions) is also an interesting direction for future research.

The connection between pattern expansion and the Hopfield model suggests potential applications in neural network theory. The Hopfield model can be viewed as a special case where patterns are explicitly stored, while pattern expansion provides a way to extract ``effective patterns'' from arbitrary spin-glass systems. This perspective might be useful for understanding how neural networks encode and retrieve information, or for developing new algorithms for pattern recognition and optimization.

From a theoretical perspective, pattern expansion provides a bridge between the Hopfield representation and the energy landscape structure of spin glasses. The fact that a finite number of expansion terms can provide a good approximation to EA models (as suggested by the rapid decay of residual norms) indicates that the energy landscape of short-range spin glasses might be more ``compressible'' than that of mean-field models. This compressibility could have implications for understanding why certain optimization algorithms work well for finite-dimensional systems but fail for mean-field models.

\clearpage
% Figure 1
\begin{figure*}
    \centering
    \begin{overpic}[width=0.98\linewidth]{figures_hop/hopfield2d.png}
        \put(-2,18){\fontsize{12}{16}\fontfamily{phv}\selectfont{\textbf{a}}}
    \end{overpic}\\[0.5cm]
    \begin{overpic}[width=0.985\linewidth]{figures_hop/hopfield3d.png}
        \put(-2,18){\fontsize{12}{16}\fontfamily{phv}\selectfont{\textbf{b}}}
    \end{overpic}\\[0.5cm]
    \begin{overpic}[width=0.98\linewidth]{figures_hop/hopfieldsk.png}
        \put(-2,18){\fontsize{12}{16}\fontfamily{phv}\selectfont{\textbf{c}}}
    \end{overpic}\\[1cm]
    \begin{overpic}[width=0.98\linewidth]{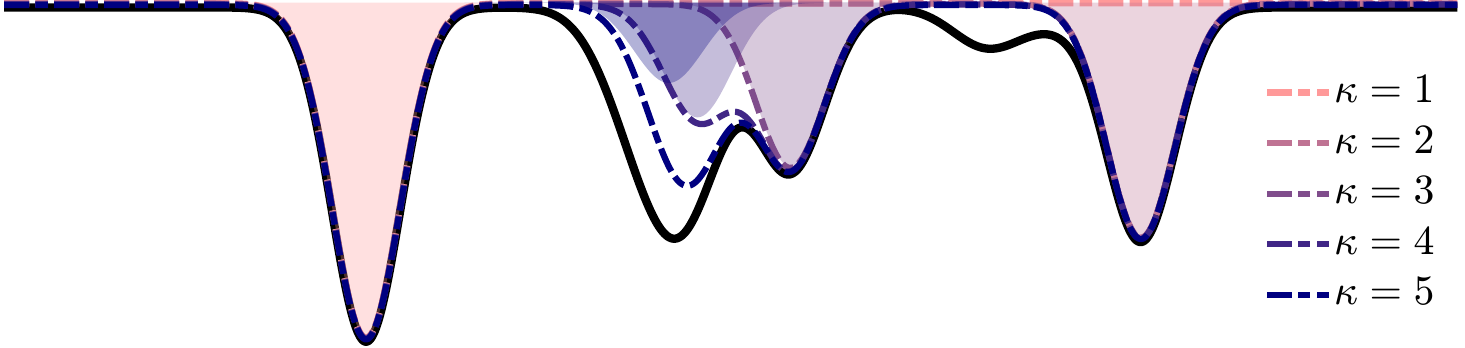}
        \put(-2,25){\fontsize{12}{16}\fontfamily{phv}\selectfont{\textbf{d}}}
    \end{overpic}
\end{figure*}
\clearpage
\begin{figure*}
    \centering
    \caption{
\textbf{Pattern expansion for different spin glass systems.} Panels (a)--(c) use simplified models to demonstrate the mapping process from original spin configurations to pattern representations. Panel (d) shows how the energy function approximates the true energy landscape as the number of patterns increases: more patterns lead to finer energy descriptions, while model complexity also increases. From top to bottom in (a)--(c) are the 2D EA model (a), 3D EA model (b), and the SK model (c). After a certain number of expansion steps, it is evident that different systems exhibit distinct behaviors in their residual systems post-expansion. In the cases of 2D and 3D, the residual system often consists of several independent subconnected sections, which are typically single bonds or frustrated plaquettes composed of four bonds. However, for the SK system, the situation is quite different. The SK model actually demonstrates a self-similar behavior, whereby the residual system after each step of expansion is another SK model. This self-similarity is prominently evident in the pattern expansion process. The existence of this self-similarity may be attributed to the mean-field nature of the SK model. In Figure \ref{fig:matrix_norm_alphas}, we will provide a more detailed explanation of the presence of this self-similarity.
    }
    \label{fig:decomp_illu}
\end{figure*}

\clearpage

% Figure 2

\clearpage

% Figure 3

\clearpage

% Figure 4

\clearpage 
\clearpage
% \clearpage 
% \clearpage 
% \begin{center}
%   \textbf{\Large Supplementary Materials}  
% \end{center}
\bibliography{ref.bib}
\bibliographystyle{nature}
\clearpage

\end{document}

\maketitle
\begin{affiliations}
    \item Department of Physics, Washington University, St. Louis, MO 63160, USA
    \item Department of Biomedical Engineering, Washington University, St. Louis, MO 63160, USA
    \item Rudolf Peierls Centre for Theoretical Physics, University of Oxford, Oxford OX1 3PU, United Kingdom
    \item LPTMC, CNRS-UMR 7600, Sorbonne Universit\'e, 4 Place Jussieu, 75252 Paris cedex 05, France
    \item Institute for Theoretical Solid State Physics, IFW Dresden, Helmholtzstrasse 20, 01069 Dresden, Germany
    \item Channing Division of Network Medicine, Department of Medicine, Brigham and Women's Hospital and Harvard Medical School, Boston, Massachusetts 02115, USA
    \\ \noindent $^\ast$Correspondence should be addressed to M.S. (smutian@wustl.edu), Z.N. (zohar@wustl.edu), or Y.-Y.L. (yyl@channing.harvard.edu).
\end{affiliations}
\setcounter{section}{0}
\renewcommand{\thesection}{S\arabic{section}}
\setcounter{figure}{0}
\renewcommand{\thefigure}{S\arabic{figure}}
\setcounter{equation}{0}
\renewcommand{\theequation}{S\arabic{equation}}
\setcounter{table}{0}
\renewcommand{\thetable}{S\arabic{table}}
\setcounter{tocdepth}{2}
\clearpage
\tableofcontents
\clearpage

\section{Gauge Transformation in EA Model}\label{sec:sm_gauge}

A gauge transformation in the Ising model is an operation that simultaneously acts on both the coupling strengths $J_{ij}$ and the spins $\sigma_i$ using a gauge vector $\boldsymbol{t} = (t_1, t_2, \ldots, t_N)$, where $t_i = \pm 1$ for all $i$. Specifically, the transformation is defined as:
\begin{align}
    J_{ij} &\to J_{ij}' = J_{ij} t_i t_j, \\
    \sigma_i &\to \sigma_i' = \sigma_i t_i.
\end{align}
It is straightforward to verify that the bond energy $E_{ij} = J_{ij} \sigma_i \sigma_j$ is invariant under this gauge transformation:
\begin{equation}
    E_{ij}' = J_{ij}' \sigma_i' \sigma_j' = J_{ij} t_i t_j \cdot \sigma_i t_i \cdot \sigma_j t_j = J_{ij} \sigma_i \sigma_j = E_{ij},
\end{equation}
where the second last equality follows from the fact that $t_i^2 = 1$ for all $i$. 

This gauge transformation can be used to demonstrate that for an EA model with coupling matrix $\mathbf{J}$, the ground state is identical to that of $-\mathbf{J}$. Since gauge transformation does not change the energy of each bond and therefore does not alter the energy landscape or physical essence of the system, we only need to show that $\mathbf{J}$ and $-\mathbf{J}$ can be transformed into each other through a gauge transformation.

\textit{Proof.} For a $d$-dimensional EA model, assume the dimensions have sizes $L_1, L_2, \ldots, L_d$ and the system is placed on a lattice $[1,L_1] \times [1,L_2] \times \cdots \times [1,L_d]$. Each node can be labeled by its coordinates $\boldsymbol{x} = (x_1, x_2, \ldots, x_d)$. We apply a gauge transformation with the rule:
\begin{equation}\label{eq:lattice_gauge_transformation}
    t_{\boldsymbol{x}} = \begin{cases}
        +1, & \text{if }  \bmod(\sum\limits_{1\leq k \leq d} x_k,2) = 1\\
        -1, & \text{otherwise}.
    \end{cases}
\end{equation} 
For any bond $J_{ij}$, since the EA model ensures that $i$ and $j$ are nearest neighbors, we have $$\left|\sum_{k=1}^{d} x_k^{(i)} - \sum_{k=1}^{d} x_k^{(j)}\right| = 1,$$ meaning we necessarily have $t_i t_j = -1$. Therefore, according to the definition in Eq. (\ref{eq:lattice_gauge_transformation}), $J_{ij}' = J_{ij} t_i t_j = -J_{ij}$. This completes the proof that $\mathbf{J}$ and $-\mathbf{J}$ can be transformed into each other through a gauge transformation.

This property allows us to compute the ground state only once during the pattern expansion process (see Eq. (\ref{eq:minimize}) in the main text).

\section{Derivation of Matrix Norm and $\alpha_\kappa$ Decay}\label{sec:sm_decay}

The self-similar behavior of the SK model under pattern expansion can be characterized through the decay of the residual matrix norm and expansion coefficients. For the first step of pattern expansion on the SK model, we start with the minimization problem in Eq. (\ref{eq:minimize}). For the SK model, the number of bonds is $|\mathcal{E}| = N(N-1)/2 \simeq N^2/2$, and the initial matrix norm is $\sumij J^2_{ij} \simeq N/2$ (since each $J_{ij}$ has variance $1/N$ and there are $N^2/2$ bonds).

When $\boldsymbol{\xi}^{(1)}$ is chosen as the ground state of $\mathbf{J}$ (or $-\mathbf{J}$), we have $\left| \sum_{\langle ij \rangle} J_{ij}\Xi_{ij}^{(1)} \right|= Ne_0$, where $e_0$ is the ground state energy per spin. From Eq. (\ref{eq:minimize}), the optimal expansion coefficient is:
\begin{equation}
\left| \alpha_1 \right| = \frac{\left| \sum_{\langle ij \rangle}J_{ij}\Xi_{ij}^{(1)}\right|}  {|\mathcal{E}|} = \frac{Ne_0}{N^2/2} = \frac{2e_0}{N}.
\end{equation}

After the first expansion step, the residual matrix is $J^{(1)}_{ij} = J_{ij} - \alpha_1 \Xi_{ij}^{(1)}$. The residual matrix norm becomes:
\begin{align}
\sumij {J^{(1)}_{ij}}^2 &= \sumij (J_{ij} - \alpha_1 \Xi_{ij}^{(1)})^2 \nonumber\\
&= \sumij J^2_{ij} - 2\alpha_1 \sumij J_{ij}\Xi_{ij}^{(1)} + \alpha_1^2 \sumij (\Xi_{ij}^{(1)})^2 \nonumber\\
&= \left(1 - \frac{4e_0^2}{N}\right) \frac{N}{2} =\left(1 - \frac{4e_0^2}{N}\right) \sumij {J_{ij}}^2 .
\label{eq:residual_norm_first_step}
\end{align}

Due to the self-similarity of the SK model, this decay factor applies at each expansion step. Therefore, after $\kappa$ steps, we have:
\begin{equation}
\sumij {J^{(\kappa)}_{ij}}^2 \simeq \left(1 - \frac{4e_0^2}{N}\right)^{\kappa-1} \frac{N}{2},
\label{eq:residual_norm_decay}
\end{equation}
which suggests an exponential decay of the residual norm.

To derive the decay of $\alpha_\kappa$, we note that at each step, the expansion coefficient is determined by the residual system. If the residual system maintains self-similarity, then $\alpha_\kappa$ should scale with the square root of the residual norm. Specifically, since $\left|\alpha_\kappa \right| \propto \sqrt{\sumij {J^{(\kappa-1)}_{ij}}^2}$, we have:
\begin{equation}
|\alpha_\kappa| \simeq \left(1 - \frac{4e_0^2}{N}\right)^{\frac{\kappa-1}{2}} |\alpha_1|.
\label{eq:alpha_decay}
\end{equation}

\begin{figure}[htb]
    \centering
    \begin{overpic}[width=0.42\textwidth]{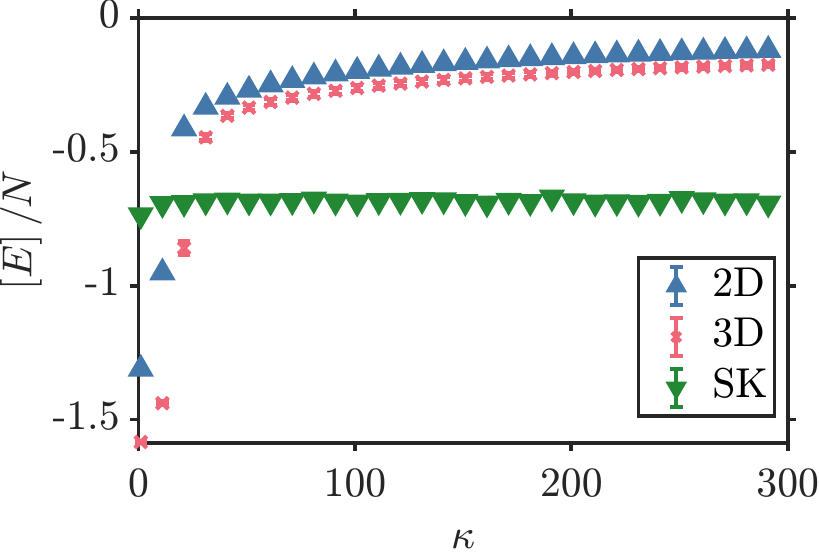}
        \put(5,70){\fontsize{12}{16}\fontfamily{phv}\selectfont{\textbf{a}}}
    \end{overpic}%
    \hspace{0.02\textwidth}%
    \begin{overpic}[width=0.42\textwidth]{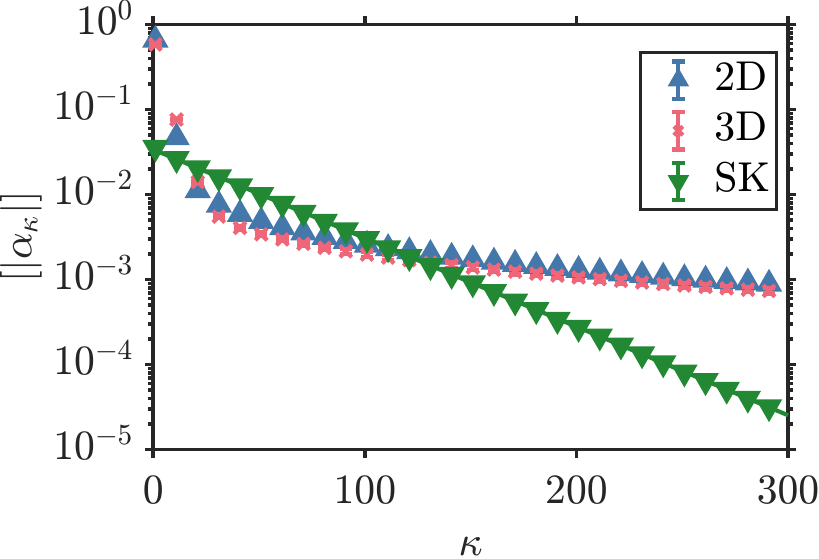}
        \put(5,70){\fontsize{12}{16}\fontfamily{phv}\selectfont{\textbf{b}}}
    \end{overpic}
    \caption{(a) Normalized residual system energy per spin $\left[ E \right] / N$ computed from the normalized residual matrix (unit standard deviation) using the ground state at each step. (b) Average absolute expansion coefficients $\left[ |\alpha_\kappa| \right]$ as a function of $\kappa$. For SK model, all quantities are normalized by $\sqrt{N}$ to account for mean-field scaling. SK shows near-exponential decay with an almost constant decay factor, while EA decays slow down as the system fractures into isolated islands. Error bars represent standard errors across multiple samples.}
    \label{fig:si_matrix_norm_alphas}
\end{figure}

This exponential decay of both the residual matrix norm and expansion coefficients reflects the self-similar nature of the SK model under pattern expansion. We verify these predictions numerically, with results shown in Fig. \ref{fig:matrix_norm_alphas} and Supplementary Fig. \ref{fig:si_matrix_norm_alphas}(b).

\section{Additional Connected-Component Analysis}
As discussed in the main text, for lattice spin-glass models the residual system after pattern expansion typically breaks into multiple isolated islands (connected components). The frequency distribution of component sizes is shown in Supplementary Fig. \ref{fig:si_normalized_energy_components}.

From another perspective, when the cumulative-system ground state is treated as an excitation of the original system, the excitation pattern can also be decomposed into disconnected islands. Some of these islands can be relatively large, reflecting the fragility of spin-glass ground states\cite{shen2023universal}. In Supplementary Fig. \ref{fig:si_kappa_cluster_counts}, we show how the number of such islands evolves as the expansion step $\kappa$ increases.

\section{Ground State Computation Methods}

All ground state computations mentioned in the main text are based on exact methods. For all EA models, we employ free boundary conditions. For 2D models, systems with free boundary conditions are planar graphs, which allows us to use the Minimum Weight Perfect Matching (MWPM) algorithm\cite{melchert2007fractal}. To reduce computational time and memory requirements, we also employ the Kasteleyn cities technique\cite{munster_cluster_2023}. For 3D models and the SK model, we use the commercial solver Gurobi\cite{gurobi_optimization_llc_gurobi_2022}, which employs branch-and-bound methods to find exact ground states.

An important consideration is the issue of numerical precision and rounding. In typical spin-glass computations, one can simply multiply the original coupling strengths $J_{ij}$ by a large factor (e.g., $10^6$) and then round to integers, which is sufficient for single ground state calculations. However, in the context of multiple pattern expansion steps, repeatedly applying such rounding would inevitably lead to error accumulation, thereby compromising the accuracy of the final computational results. To address this, we adopt a more careful approach: we multiply the original couplings by a large factor ($10^6$) at the beginning, but then maintain floating-point precision throughout all subsequent calculations. Specifically, during the expansion process, we compute the recursive relation $J^{(\kappa)}_{ij} = J^{(\kappa-1)}_{ij}-\alpha_{\kappa}\Xi^{(\kappa)}_{ij}$ (see the main text) using floating-point arithmetic. Rounding is only performed at the final stage when outputting bond weights. This strategy ensures that numerical errors do not accumulate during the iterative expansion process, which is crucial for accurately tracking the evolution of the residual system and computing low-energy excitations.

\section{Validity of Pattern Expansion with Approximate Ground-State Solvers}
For large systems (especially in 3D), exact ground-state solvers become prohibitively expensive when many disorder realizations are required (e.g., $L>10$ and even $L>20$). In this regime, approximate methods are necessary. Typical choices include simulated annealing\cite{bertsimas1993simulated,isakov2015optimised}, parallel tempering\cite{roma2009ground}, population annealing\cite{amey2018analysis,barzegar2018optimization}, and, more recently, learning-based solvers\cite{RL_Local,mills2020finding}.

Here we use simulated annealing (SA) as the approximate ground-state engine, and test whether the resulting expansion remains consistent with the expansion based on exact ground states.

\paragraph{Simulated annealing protocol used in this work.}
Given couplings $J_{ij}$ and a spin configuration $\boldsymbol{\sigma}\in\{\pm1\}^N$, the Hamiltonian is
\begin{equation}
\mathcal{H}(\boldsymbol{\sigma})=-\sum_{\langle i j \rangle}J_{ij}\sigma_i\sigma_j.
\end{equation}
At each Monte Carlo proposal, we randomly choose a site $i$ and propose $\sigma_i\to -\sigma_i$. The energy increment is computed as
\begin{equation}
\Delta E = 2 \sigma_i \sum_j J_{ij}\sigma_j,
\end{equation}
and the move is accepted with the Metropolis rule:
\begin{equation}
P_{\mathrm{acc}}=\min\!\left(1,e^{-\beta \Delta E}\right).
\end{equation}
We use a linear inverse-temperature schedule $\beta\in[\beta_{\min},\beta_{\max}]$, with defaults $\beta_{\min}=10^{-3}$, $\beta_{\max}=5$, and $N_T=100$ temperature points. For each temperature, we run $n_{\mathrm{sweep}}=50$ sweeps, and each sweep contains $N$ single-spin proposals. The initial state is a random $\boldsymbol{\sigma}\in\{\pm1\}^N$, and the random seed is shuffled by default (or fixed when specified).

\begin{figure}[htb]
    \centering
    \begin{overpic}[width=0.44\textwidth]{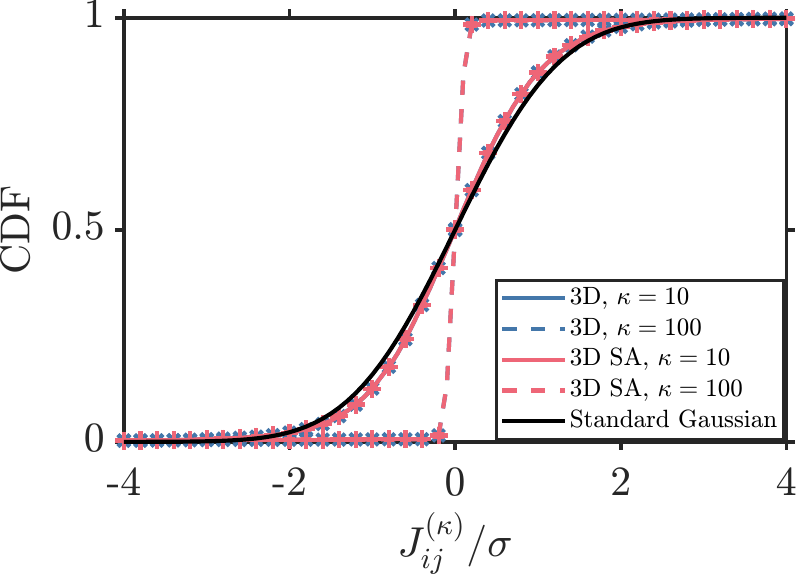}
        \put(4,74){\fontsize{12}{16}\fontfamily{phv}\selectfont{\textbf{a}}}
    \end{overpic}%
    \hspace{0.02\textwidth}%
    \begin{overpic}[width=0.46\textwidth]{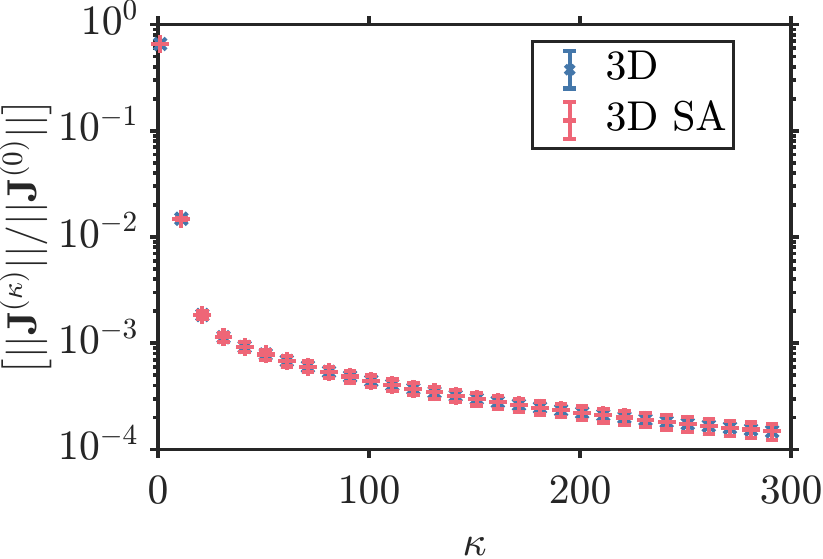}
        \put(8,70){\fontsize{12}{16}\fontfamily{phv}\selectfont{\textbf{b}}}
    \end{overpic}
    \begin{overpic}[width=0.46\textwidth]{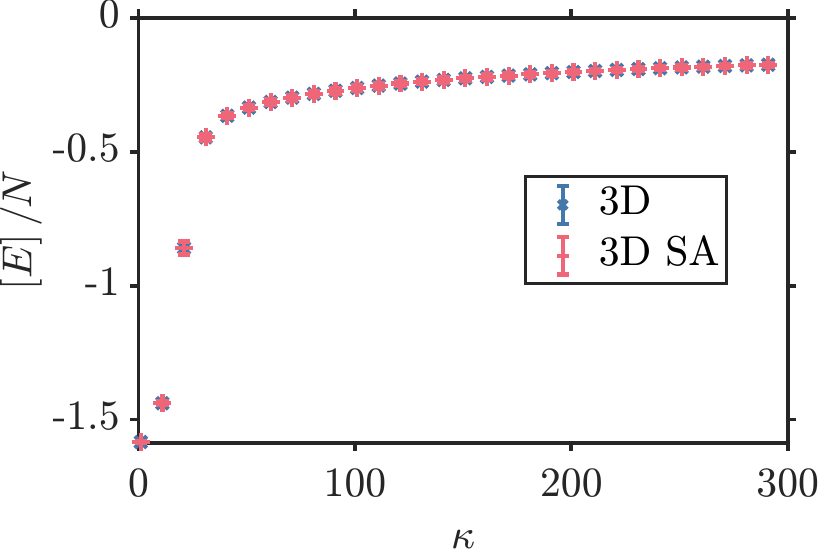}
        \put(5,70){\fontsize{12}{16}\fontfamily{phv}\selectfont{\textbf{c}}}
    \end{overpic}%
    \hspace{0.02\textwidth}%
    \begin{overpic}[width=0.46\textwidth]{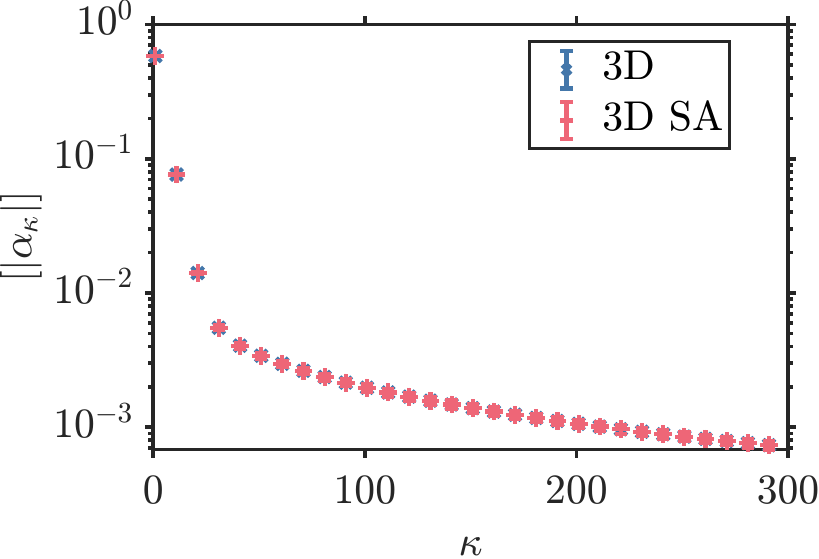}
        \put(5,70){\fontsize{12}{16}\fontfamily{phv}\selectfont{\textbf{d}}}
    \end{overpic}
    \caption{Comparison between exact (Gurobi) and simulated-annealing (SA) based pattern expansion. (a) CDF of normalized residual couplings $J_{ij}^{(\kappa)}/\sigma^{(\kappa)}$ at representative expansion steps. (b) Normalized residual norm $\avg{\|\mathbf{J}_{\kappa}\|/\|\mathbf{J}_0\|}$ as a function of $\kappa$. (c) Normalized residual-system energy density $\avg{E}/N$. (d) Average absolute expansion coefficients $\avg{|\alpha_\kappa|}$. Across all observables, SA closely tracks the exact-solver baseline, indicating that the expansion pipeline is robust to the choice of ground-state solver.}
    \label{fig:si_sa_comparison}
\end{figure}

To validate the expansion with an approximate ground-state solver, we compare the residual-system statistics obtained from SA against those from the exact solver (Gurobi) across the same disorder samples and expansion steps, see Fig. \ref{fig:si_sa_comparison}.

\section{Population Annealing with Thermal Boundary Conditions for the 3D Edwards--Anderson Spin Glass}
\label{sec:supp_pa}

We consider the three-dimensional Edwards--Anderson (EA) Ising spin glass on a cubic lattice of linear size $L$ with $N=L^3$ spins. The population annealing algorithm and the use of thermal boundary conditions in this section follow mainly Refs.\cite{wangPopulationAnnealingMonte2015,wangPopulationAnnealingTheory2015}.

\subsection{Reference simulation parameters}

The large-scale PA benchmarks in Ref.\cite{wangPopulationAnnealingMonte2015} use, for each linear size $L$, a population size $R$, a lowest temperature $T_0$ (so $\beta_{\max}=1/T_0$), $N_T$ inverse-temperature points evenly spaced in $\beta\in[0,\beta_{\max}]$, and $N_S$ Metropolis sweeps after each resampling step. Table \ref{tab:pa_params_wang} reproduces Table 2.1 of that reference (columns $R$, $T_0$, $N_T$; $N_S=10$ is fixed across entries there). The same numbers are implemented in our code as the optional ``published'' schedule keyed by $L$; we have checked them entrywise against Ref.\cite{wangPopulationAnnealingMonte2015}. Independently of $L$, overlap moments use $n_{\mathrm{rep}}=3$ independent random pairings per temperature (Sec. \ref{sec:pa_overlap})---a variance-reduction device not listed in the same table.

\subsection{Thermal Boundary Conditions}

To reduce finite-size effects, we employ thermal boundary conditions (TBC), in which all $2^d$ combinations of periodic and antiperiodic boundary conditions are included in the ensemble. In $d=3$, this gives $2^3=8$ boundary sectors labeled by
\begin{equation}
    \mathbf{b} = (b_x, b_y, b_z), \quad b_\mu \in \{+1,-1\},
\end{equation}
where $+1$ ($-1$) denotes periodic (antiperiodic) boundary conditions along $\mu$.

Symmetric sector matrices $\mathbf{J}_{\mathbf{b}}$ are built from a single periodic reference (all bonds periodic) by assigning signs on bonds that cross the periodic boundaries; each replica carries one sector label $\mathbf{b}$. The partition function and thermal weights $w_{\mathbf{b}} = Z_{\mathbf{b}}/\sum_{\mathbf{b}'}Z_{\mathbf{b}'}$ are the usual TBC ensemble definitions; population annealing does not evaluate these weights explicitly, but samples sectors with frequencies that track the Boltzmann weights during annealing.

\subsection{Hamiltonian and energy}

Each replica uses a symmetric sparse interaction matrix $\mathbf{J}_{\mathbf{b}}$ of size $N\times N$. With spins $\sigma_i\in\{\pm 1\}$,
\begin{equation}
    H_{\mathbf{b}}(\boldsymbol{\sigma}) = -\frac{1}{2}\,\boldsymbol{\sigma}^{\mathsf T}\mathbf{J}_{\mathbf{b}}\boldsymbol{\sigma},
\end{equation}
and single-spin Metropolis flips use $\Delta E = 2\sigma_i\,(\mathbf{J}_{\mathbf{b}}\boldsymbol{\sigma})_i$, consistent with the usual Edwards--Anderson symmetric-matrix convention.

\subsection{Population Annealing Monte Carlo}

Population annealing (PA) evolves $R$ replicas along an inverse-temperature ladder from $\beta=0$ (infinite temperature) toward a target low temperature set by a minimum temperature $T_0$ (maximum inverse temperature $\beta_{\max}=1/T_0$).

\paragraph{Initialization.}
At $\beta=0$, each replica draws an i.i.d.\ random spin configuration and a boundary label $\mathbf{b}$ drawn uniformly from the eight sectors.

\paragraph{Annealing schedule.}
We set $\beta_{\max}=1/T_0$ and use $N_T$ \emph{uniformly spaced} inverse temperatures from $0$ to $\beta_{\max}$. For $N_T\ge 2$ this is
\begin{equation}
    \beta_k = \frac{k-1}{N_T-1}\,\beta_{\max},\qquad k=1,\ldots,N_T,\qquad \beta_{\max}=\frac{1}{T_0},
\end{equation}
so $\beta_1=0$ and $\beta_{N_T}=\beta_{\max}$. The corresponding temperatures are $T_k=1/\beta_k$ for $\beta_k>0$ and $T=\infty$ at $\beta=0$.

\paragraph{Resampling.}
When advancing $\beta\to\beta'=\beta+\Delta\beta$, replica $i$ with total energy $E_i$ (for its own $\mathbf{b}^{(i)}$) receives a Boltzmann factor
\begin{equation}
    \tau_i = \exp\bigl[-\Delta\beta\,E_i\bigr].
\end{equation}
Expected copy numbers are proportional to $\tau_i/\overline{\tau}$ with $\overline{\tau}=R^{-1}\sum_j \tau_j$, and the population is redrawn with \emph{nearest-integer} resampling so that the total number of replicas remains (approximately) $R$.

\paragraph{Monte Carlo updates.}
At the first ladder point ($\beta=0$), overlaps are measured on the initial random population without preceding Metropolis sweeps: at infinite temperature the Boltzmann factor does not distinguish spin configurations, so additional local moves are omitted there. For each subsequent step $k=2,\ldots,N_T$, we first apply the resampling rule at fixed $(\beta_{k-1},\beta_k)$, then perform $N_S$ full sweeps of single-spin Metropolis updates at $\beta_k$ under $\mathbf{J}_{\mathbf{b}}$, and then record observables.

\paragraph{Boundary condition propagation.}
The label $\mathbf{b}$ is carried with each replica and copied on resampling; the empirical sector fractions $R_{\mathbf{b}}/R$ (used for monitoring) need not coincide with the exact $w_{\mathbf{b}}$ at finite $R$ and finite annealing length.

\subsection{Overlap and Moments}
\label{sec:pa_overlap}

For a pair of replicas $(a,b)$ with spin configurations $\boldsymbol{\sigma}^{(a)}$ and $\boldsymbol{\sigma}^{(b)}$, the overlap is
\begin{equation}
    q^{(ab)} = \frac{1}{N}\sum_{i=1}^{N} \sigma_i^{(a)}\sigma_i^{(b)}.
\end{equation}

At each temperature we draw several independent \emph{random disjoint pairings} of the replica list; each pairing is then adjusted by local partner exchanges so that every retained link joins two replicas from different PA lineages (``non-incest'' pairs with respect to an integer family tag). Pairs are \emph{not} restricted to the same $\mathbf{b}$, so the estimate targets the full TBC ensemble (including pairs that may lie in different boundary sectors). The draw-and-repair step is repeated $n_{\mathrm{rep}}$ times (we use $n_{\mathrm{rep}}=3$) to reduce the variance of the empirical moments. The second and fourth moments are estimated as \emph{sample means} over all collected links,
\begin{equation}
    \langle q^2\rangle = \overline{(q^{(ab)})^2},\qquad
    \langle q^4\rangle = \overline{(q^{(ab)})^4},
    \label{eq:pa_q_moments}
\end{equation}
where the overline denotes an average over pair samples.

\subsection{Binder Cumulant}

We characterize overlap fluctuations through low moments rather than a binned estimate of the full distribution $P(q)$. From the moments in Eq. \eqref{eq:pa_q_moments},
\begin{equation}
    g = \frac{1}{2}\left(3 - \frac{\langle q^4\rangle}{\langle q^2\rangle^2}\right),
\end{equation}
with a small numerical floor on $\langle q^2\rangle^2$ in the denominator to avoid division by zero. The cumulant is formed from these \emph{global} moment estimates (not from averaging Binder values computed separately within each boundary sector).

\subsection{Disorder averaging, Hop expansion, and figures}

For each disorder realization of the bare matrix $\mathbf{J}_0$, we run one PA trajectory and obtain $g(T)$ (and related quantities). When coupling matrices from the Hop expansion are used, the bare $\mathbf{J}_0$ and the cumulative matrices at expansion steps $\kappa$ may be rescaled by a common factor before PA so that energies lie in a stable numerical range; each realization is still one PA run.

When aggregating many disorder realizations, at each temperature we take the mean of $g$ across samples and, for figures, the standard error of the mean, $\mathrm{SEM}=\sigma_g/\sqrt{n_{\mathrm{eff}}}$, where $\sigma_g$ is the sample standard deviation across realizations and $n_{\mathrm{eff}}$ the number of valid samples at that $T$.

\section{Explanation of Frustrated Plaquettes Density Decrease at $\kappa=4$}

\begin{figure}[htb]
    \centering
    \includegraphics[width=0.8\textwidth]{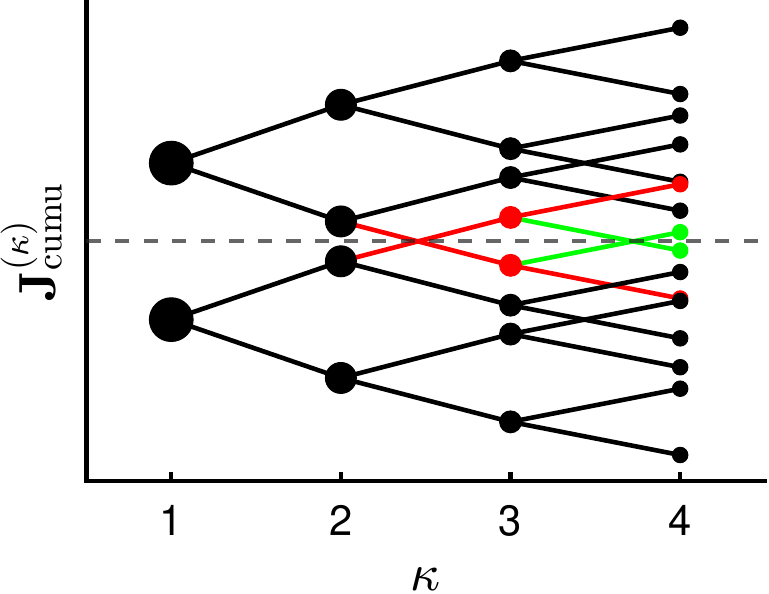}
    \caption{Evolution of cumulative coupling values $J_{\text{cumu}}^{(\kappa)}$ as a function of expansion order $\kappa$ from $\kappa=1$ to $\kappa=4$. The horizontal dashed line represents the zero line. At each expansion step, the number of distinct possible values doubles, reflecting the $2^\kappa$ combinations of $\pm\alpha_\mu$ coefficients. The red nodes and red path illustrate sequences where the sign flips at $\kappa=3$ (crossing the zero line) and then also at $\kappa=4$, demonstrating the sign reversion that leads to the decrease in frustrated plaquettes density. The green nodes and green path show sequences where bonds that had already flipped sign but then flipped back. The $\alpha_\mu$ values used in this calculation are obtained from a real 2D EA system with $L=256$.}
    \label{fig:division}
\end{figure}

At $\kappa=4$ (see Fig. \ref{fig:frustration} in the main text), we observe a somewhat unusual phenomenon: the density of frustrated plaquettes $\rho$ suddenly decreases. The explanation for this is straightforward and is illustrated in Fig. \ref{fig:division}.

For any expansion order $\kappa$, the cumulative coupling matrix $\mathbf{J}_{\text{cumu}}^{(\kappa)}$ has a limited number of distinct nonzero element types. For example, at $\kappa=3$, the elements take the form $k_1\alpha_1 + k_2\alpha_2 + k_3\alpha_3$, where $k_1, k_2, k_3 \in \{\pm 1\}$ and $\alpha_1, \alpha_2, \alpha_3$ are the expansion coefficients of the Mattis interactions. In general, there are $2^\kappa$ distinct types of nonzero elements. The distinct types of nonzero elements displayed in Fig. \ref{fig:division} are computed from the $\alpha_\mu$ values obtained from a real 2D EA system with $L=256$.

What determines $\rho$ is actually the sign flip behavior of these nonzero elements. At $\kappa=1$, there is no frustration because, under gauge transformation, the system is equivalent to a ferromagnetic system. At $\kappa=2$, since $|\alpha_2| < |\alpha_1|$, the perturbation from the second-order term cannot change the sign of any bond's weight. However, at $\kappa=3$, the situation is quite different. As can be seen in Fig. \ref{fig:division} (red nodes), combinations that change the sign of bond weights appear, specifically $\pm(\alpha_1 - \alpha_2 - \alpha_3)$. This means that some bonds change sign, leading to the emergence of frustration. At $\kappa=4$, the combination $\pm(\alpha_1 - \alpha_2 - \alpha_3 + \alpha_4)$ flips the sign back, while $\pm(\alpha_1 - \alpha_2 + \alpha_3 - \alpha_4)$ necessarily has the same sign as $\pm\alpha_1$ (since the $\alpha_\mu$ coefficients decay with $\mu$). In summary, from $\kappa=3$ to $\kappa=4$, bonds that did not change sign remain unchanged, but some bonds that had changed sign at $\kappa=3$ revert back to their original sign, thereby reducing the density of frustrated plaquettes $\rho$.

\clearpage

\begin{table}
    \centering
    \begin{tabular}{ccccc}
        \hline
        $L$ & $R$ & $T_0$ & $N_T$ & $N_S$ \\
        \hline
        $4$  & $5\times 10^{4}$  & $0.20$  & $101$ & $10$ \\
        $6$  & $2\times 10^{5}$  & $0.20$  & $101$ & $10$ \\
        $8$  & $5\times 10^{5}$  & $0.20$  & $201$ & $10$ \\
        $10$ & $10^{6}$          & $0.20$  & $301$ & $10$ \\
        $12$ & $10^{6}$          & $0.333$ & $301$ & $10$ \\
        $14$ & $3\times 10^{6}$  & $0.333$ & $401$ & $10$ \\
        \hline
    \end{tabular}
    \caption{Population annealing parameters for the three-dimensional EA model with thermal boundary conditions, as in Table 2.1 of Ref.\cite{wangPopulationAnnealingMonte2015}. $R$ is the number of replicas, $T_0$ the lowest simulated temperature, $N_T$ the number of $\beta$ points (including $\beta=0$), and $N_S$ the number of full Metropolis sweeps per replica after each resampling step.}
    \label{tab:pa_params_wang}
\end{table}

\clearpage

\begin{figure}[htb]
    \centering
    \begin{overpic}[width=0.42\textwidth]{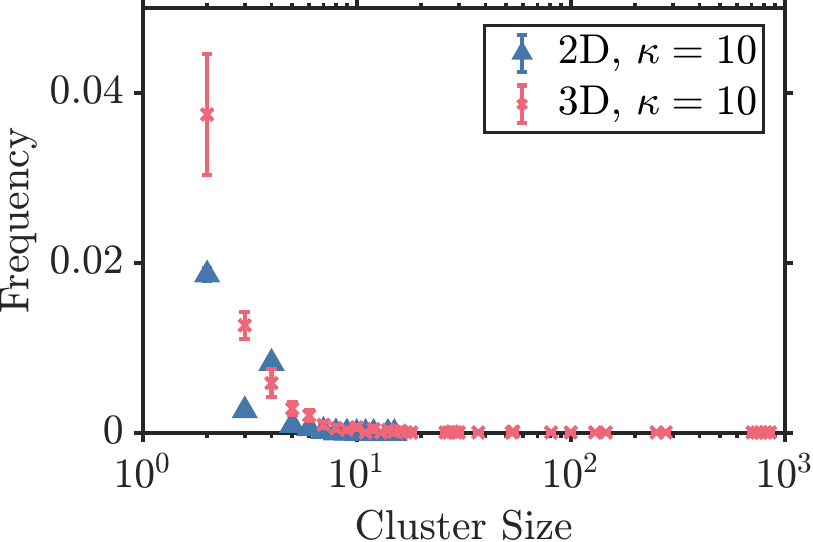}
        \put(7,69){\fontsize{12}{16}\fontfamily{phv}\selectfont{\textbf{a}}}
    \end{overpic}%
    \hspace{0.02\textwidth}%
    \begin{overpic}[width=0.41\textwidth]{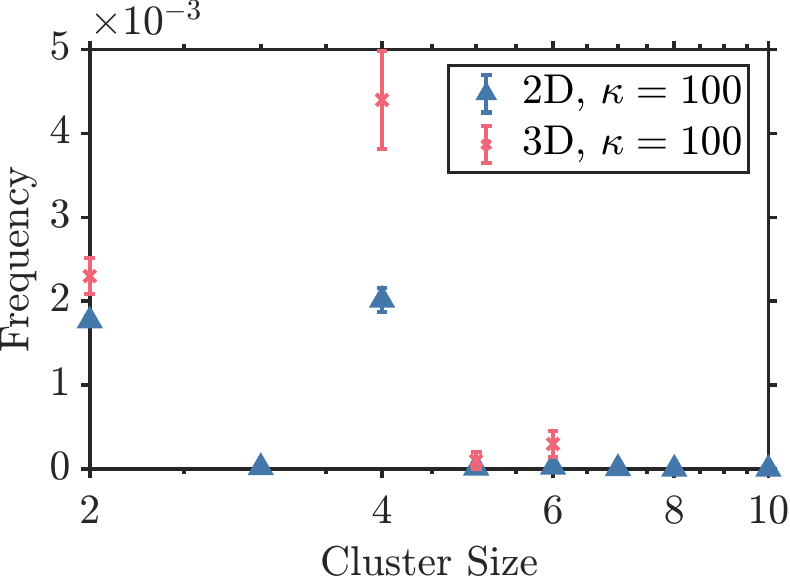}
        \put(1,71){\fontsize{12}{16}\fontfamily{phv}\selectfont{\textbf{b}}}
    \end{overpic}
    \caption{Frequency distribution of connected component sizes in the residual system for 2D (triangles) and 3D (crosses) EA models at $\kappa = 10$ (a) and $\kappa = 100$ (b). Components are identified by selecting edges with $|J_{ij}^{(\kappa)}| > \rho\max(|J_{ij}^{(\kappa)}|)$ in the normalized residual matrix, where we set $\rho=0.1$. The distribution shows that EA models break into isolated subconnected sections (small clusters). At $\kappa=100$, in all computed results, no connected components larger than 10 exist.}
    \label{fig:si_normalized_energy_components}
\end{figure}

\begin{figure}[htb]
    \centering
    \begin{overpic}[width=0.42\textwidth]{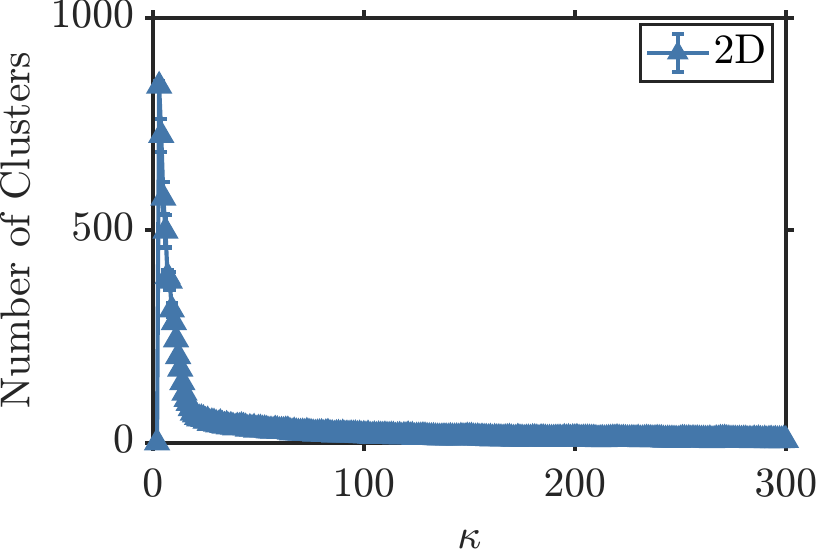}
        \put(3,71){\fontsize{12}{16}\fontfamily{phv}\selectfont{\textbf{a}}}
    \end{overpic}%
    \hspace{0.02\textwidth}%
    \begin{overpic}[width=0.42\textwidth]{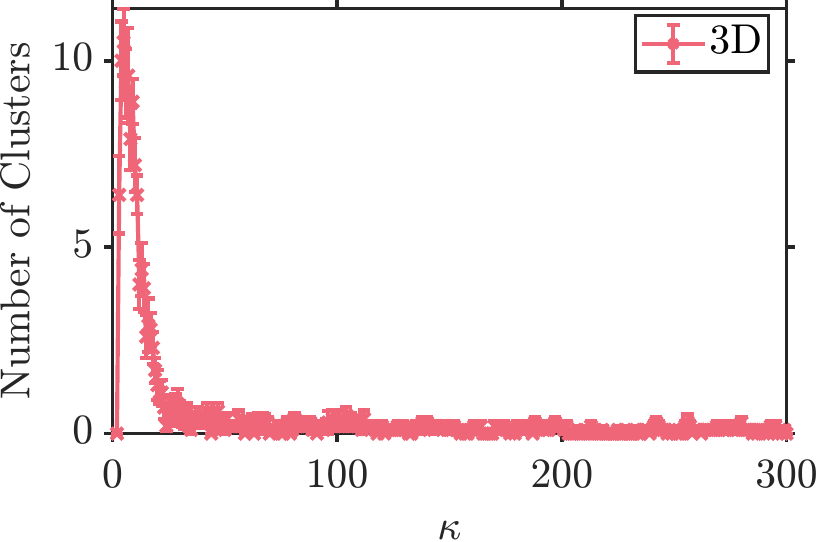}
        \put(1,70){\fontsize{12}{16}\fontfamily{phv}\selectfont{\textbf{b}}}
    \end{overpic}
    \caption{Number of independent excitation clusters (connected components) as a function of expansion step $\kappa$ for 2D (a) and 3D (b) EA models. Each data point represents the average number of clusters across all samples, with error bars showing the standard error. The number of independent excitations gradually decreases with $\kappa$.}
    \label{fig:si_kappa_cluster_counts}
\end{figure}

\clearpage

\bibliography{ref.bib}
\bibliographystyle{nature}